# Darknet Traffic Big-Data Analysis and Network Management to Real-Time Automating the Malicious Intent Detection Process by a Weight Agnostic Neural Networks Framework

**Konstantinos Demertzis*[1,5], Konstantinos Tsiknas[2], Dimitrios Takezis[3], Charalabos Skianis[4] and Lazaros Iliadis[5]**

[1] Laboratory of Complex Systems, Department of Physics, Faculty of Sciences, International Hellenic University, Kavala Campus, St. Loukas, 65404, Greece; kdemertzis@teiemt.gr
[2] Department of Electrical and Computer Engineering, Democritus University of Thrace, Vas. Sofias 12,, 67100, Xanthi, Greece; ktsiknas@ee.duth.gr
[3] Hellenic National Defence General Staff, Stratopedo Papagou, Mesogeion 227-231, 15561 Athens, Greece; d.taketzis@hndgs.mil.gr
[4] University of Aegean, 83200, Karlovassi, Samos, Greece; cskianis@aegean.gr
[5] School of Engineering, Department of Civil Engineering, Faculty of Mathematics Programming and General courses, Democritus University of Thrace, Kimmeria, Xanthi, Greece
* Correspondence: kdemertzis@teiemt.gr;

**Abstract:** Attackers are perpetually modifying their tactics to avoid detection and frequently leverage legitimate credentials with trusted tools already deployed in a network environment, making it difficult for organizations to proactively identify critical security risks. Network traffic analysis products have emerged in response to attackers' relentless innovation, offering organizations a realistic path forward for combatting creative attackers. Additionally, thanks to the widespread adoption of cloud computing, Device Operators (DevOps) processes, and the Internet of Things (IoT), maintaining effective network visibility has become a highly complex and overwhelming process. What makes network traffic analysis technology particularly meaningful is its ability to combine its core capabilities to deliver malicious intent detection. In this paper, we propose a novel darknet traffic analysis and network management framework to real-time automating the malicious intent detection process, using a weight agnostic neural networks architecture. It is an effective and accurate computational intelligent forensics tool for network traffic analysis, the demystification of malware traffic, and encrypted traffic identification in real-time. Based on Weight Agnostic Neural Networks (WANNs) methodology, we propose an automated searching neural net architectures strategy that can perform various tasks such as identify zero-day attacks. By automating the malicious intent detection process from the darknet, the advanced proposed solution is reducing the skills and effort barrier that prevents many organizations from effectively protecting their most critical assets.

---

## 1. Introduction

Interconnected heterogeneous information systems [1] exchange huge amounts of data per unit of time. This information consists of data at rest and data in motion. In the continuous flow model, the data arrives in successive streams in a continuous manner, resulting in not being accessible by the storage mediums, either temporary or permanently. Flow data is usually large in size, difficult to be processed in real-time, and when processed, it is either destroyed or archived, and is very difficult to be recovered again, because the system's memory is typically very small.

The analysis, monitoring, and categorization of the Internet network traffic [2] is one of the most important tasks and characterised as a specialised solution and a valuable tool that can be used not only to effectively deal with the design, management, and monitoring of the critical infrastructure of the system, but also for the monitoring of attacks and the study of cybercrime [3].

The information exchanged can be requests, responses, or control data, fragmented in the form of network packets. When looking at individual network packets, it is extremely difficult to draw conclusions and exclude safe conclusions, as the information transmitted between devices on the network is fragmented into a number of packets, which are interconnected, containing all the information. This arbitrary and occasional nature of the collection of network traffic, while providing some information for drawing statistical conclusions, makes the use of typical mathematical analysis methods a rather difficult task which favours the network traffic modeling approach [4].

Many organizations, in their efforts to improve and enhance their security, collect as much web traffic data as possible, analyze it, by correlating it with the services they represent, and compare it with historical log files in order to optimize their decision-making process. By analyzing network traffic, safe conclusions can be drawn about the network, the users, and the total data usage, making it possible to model traffic in order to optimize network resources according to the monitoring needs and the control for legal and security issues [5], [6]. More specifically, in cybersecurity, traffic analysis can be applied to secure services, guarantee critical data delivery, identify random sources of problems, adapt and optimize intrusion prevention and detection solutions, identify cybercriminals, and validate forensic data [7]. The two basic methods of network traffic analysis are:

1. Conventional Packet Filtering (CPI). The basic, conventional way of analyzing network traffic, which is also called shallow analysis, is related to the encapsulation process by which information is exchanged on a network. This process is a strict way of separating the data and metadata contained in the network information. With this method,a complete examination of the web traffic can be performed, without any problem regarding privacy policy violations. However, the problem with this method is that only the headers of the network packets (metadata) can be extracted, while the actual traffic data (payload of the packet), cannot be extracted and analyzed due to the large size, which is also required and necessary. The result is, that this method is not appropriate for environments, where the security of the transmitted information is an important operating factor.

2. Deep Packet Inspection (DPI). In-depth packet inspection, also known as full packet inspection, is a type of filtering that evaluates the data segment (packet payload) along with the packet header that is transmitted, detecting any non-compliance, or redirection of the packet to another destination. In short, in-depth package inspection can detect, categorize, block, or recall packages that have a specific payload or data that is not detected, categorized, blocked, or redirected by the Conventional Packet Filtering. The most basic technique of deep packet inspection for the purpose of detecting digital attacks involves pattern or signature matching. This method performs packet analysis to detect patterns or signatures mapped from a database of known network attacks. The disadvantage of this approach is that it is effective only for known attacks, but not for attacks that have not yet been discovered.

The major weaknesses associated with traffic packet analysis technologies are the following [8]:

1. While the techniques are very effective, especially the DPI method in preventing Denial of Service (DoS) - Distributed DoS(DDoS) attacks, buffer overflow attacks, and specific types of malwares, they can also be used to create similar attacks from the adversary side, depending on their mode of operation.

2. They add complexity to the operation of active network security methods and make them extremely difficult to manage. In additionThey increase the requirements for computing resources and introduce significant delays in online transactions, especially in encrypted traffic, since the latter requires the reconstruction of messages and entities at higher levels.

3. Although there are many possible uses, an adverse situation is related to the ease with which someone can identify the recipient, or the sender of the content they are analyzing, raising privacy concerns.
4. They do not offer protection against zero-day attacks.

The ever-increasing need for an organization to manage security incidents requires specialized analysis services, in order to fully understand the network environment and potential threats. This information, combined with cyber threat intelligence from the global threat landscape, allows for an informed and targeted response to cyber-related incidents [9].

In essence, the information ecosystem and the importance of its applications require the creation of a cybersecurity environment with fully automated solutions. These solutions include real-time incident handling, analysis, and other security information to identify known and unknown threats and reduce the risk for the critical data through a scalable troubleshooting and logging approach [10], [11].

## 2. Network Management

In this context, the specialized network traffic analysis services that can be used to secure network applications and industrial confidentiality[12]are presented below.

### 2.1 Traffic Classification

Network traffic categorization, is the automated process that classifies network traffic into categories, based on evaluation criteria. Each resulting category can be treated differently in order to differentiate the services provided to the user and to implement the appropriate security policies. It is considered as a very important process in network management, as it is related with all applications and services running online, and affects the ways and policies applied for securing and evaluating the services in question. It also influences the methodologies associated with the application rules defining the Quality of Service (QoS) of the network in use [13].

There are two basic approaches to classify network traffic:

1. Classification based on the network packet payload method, by which the packets are sorted by protocols and domains of that payload, such as Layer 2, for instance Medium Access Control (MAC) address, Layer 3, for instance Internet Protocol (IP)IP address, and Layer 4 (source or destination ports, or both of them).
2. Classification based on a statistical analysis of traffic behavior, such as arrival time, waiting time between network packets, session time, etc.

The network traffic classification is accomplished by specialised software capable of analysing the contents of the packets, detecting the absence of a packet and determining the basic relations that some packet metrics have, such as the Transmission Control Protocol (TCP) sequence and confirmation numbers, etc. This weakness is due to the fact that these tools provide only top-level information, without the corresponding ones coming from the metadata (packet headers) that can identify events and therefore deal effectively with network problems.

For example, firewalls, and especially firewalls that incorporate Intrusion Detection System (IDS) and Intrusion Prevention System (IPS) capabilities into their operation [14], can be considered as network traffic classification software, they provide a variety of services such as Network Address Translation (NAT), Port Address Translation (PAT), Virtual Private Network (VPN) and Egress Filtering for traffic classification that does not comply with network security policies.

Despite their potential, firewalls must also provide adequate capabilities for detecting and preventing intrusions, such as malware detection, instant messaging services analysis, and SSL (Secure Socket Layer) session inspection , in order to be considered an adequate and robust security solution [15].

Finally, an important disadvantage that further complicates the procedures for categorizing network traffic, is that the processes in question are very demanding in computing resources, as their secure categorization requires the decryption of a session eg. SSL, and then the reconstruction of the session packets.

## 2.2 Demystifying Malicious Traffic

The application of malicious traffic detection refers to the capture, recording, and analysis of data in the form of packets, which are related to attacks on the network or a computer, attempts to breach security, or generally unusual activity. Given the multitude of modern attacks that have the ability to cover their tracks, or provide distorted or insufficient data to conduct research, identifying malicious traffic is an extremely arduous, difficult, and particularly important task of cybersecurity [16].In essence, this is the main method of preventing and detecting cyber attacks, as it detects command and control communications of malware. It should also be noted that it is the most important process in assessing the behavior of malicious software, revealing the purpose they serve and the potential damage they can cause.

Malware, in particular, is a type of software used to gain unauthorized access to a computer, collect sensitive information, delete private data, or even shut it down. It can appear in the form of code, scripts, active content, or any other programming methodology. Recent malware developments have the potential to remain hidden during system infection and operation. They prevent analysis and removal using a variety of techniques, such as obscure filenames, modify file attributes or properties, or hiding and running legal processes or programs. Malicious software may also attempt to compromise security software by hiding running processes, network connections, and strings of a program or service with malicious Uniform Resource Locators (URLs) or registry keys.

The most popular types of malware seek to retrieve and maintain communication on a regular basis with Command and Control servers (C&C servers) which are under the control of a malicious user, so that they can collect, transfer information, and upgrades to the infected devices (bots) [17], [18]. This communication is usually done using hardcoded addresses, or a pool of addresses controlled by the creator of the malware or network.Modern programming techniques enable malware developers to use thousands of alternating IP addresses to communicate with C&C servers. These IPs are relatively simple for network engineers or security analysts to detect, block and blacklist.

## 2.3 Encrypted Traffic Recognition

The categorization of encrypted traffic is considered as one of the most important processes of active security and protection of web applications, although it is one of the most serious challenges of modern computing.In particular, encryption is known to provide users with confidentiality and privacy protection, by hiding the flow of data and preventing their identification. At the same time, it introduces difficulties in  the identification and categorisation of the important business applications and prevents the immediate prioritisation of high-importance processes from achieving an optimal performance. In addition, it is hardly managable in cases of network congestion and does not easily allow for the  restriction of unauthorized applications to access  enterprise networks such as pornography, gambling, etc. [12][15].

Similarly, as most organizations use encryption as the primary method of securing the information they exchange, the rapid increase in the use of this service has changed the threat landscape, as cybercriminals use the same method to avoid detection of their malicious activities. A typical example is the use of the most important modern anonymity encryption service, such as Tor (The Onion Router), which is the most famous way of communicating with the latest, advanced generations of malware, such as Darknet communications are based on the creation of an encrypted communication channel based on chaotic architectures, in order to alter the traces, distort the elements that identify an attack and ultimately increase the complexity of malicious networks [20].

## 3. Tor hidden services on Darknet

The visible layer of the web that users can access through search engines is only a small part of the internet. The part of the internet that is not accessible by search engines is also known as the Deep Web. Darknet is a subset of the Deep Web in the sense that it is also undetectable by search engines, but it also has one characteristic [21]. Anonymous communication networks - commonly referred to as Darknets - are becoming increasingly popular with criminals for trafficking drugs, weapons, fake IDs, theft, and child pornography. In fact, a completely illegal online marketplace has been developed which provides all kinds of illegal services which is facilitated by the use of cryptocurrencies and covered by anonymity. To achieve this anonymity, communication is done using special networks that pronounce anonymity such as Tor and I2P (Invisible Internet Project) [22].

Tor software routes web traffic through a peer-to-peer network. It is the implementation of onion routing, in which multi-layer encryption is being used, it ensures perfect forward secrecy between the nodes and the hidden services of Tor, while at the same time it routinely communicates through Tor nodes (consensus) operated by volunteers around the world. This is achieved by using virtual circuits or overlays, which change at regular intervals. Although the Tor network operates at OSI Level 4 (Transport Layer), the onion proxy software displays to clients the Socket Secure (SOCKS) interface that operates at Level 5 (Session layer). Also, in this network, there is the continuous redirection of requests between the retransmission nodes (entry guards, middle relays, and exit relays), with the sender and recipient addresses as well as the information being encrypted, so that no one at any point along the communication channel, can directly decrypt the information or identify both ends [23].

The Tor network not only provides encryption, but is also designed to emulate the normal traffic of the HTTPS protocol, making the detection of Tor channels an extremely complex and specialized process, even for experienced network engineers or analysts. Specifically, the Tor network can use the TCP port 443, which is also used by Hypertext Transfer Protocol Secure (HTTPS), so monitoring and identifying a session solely by the port is not a reliable method of determining this type of traffic [24].

A successful method for detecting Tor traffic involves statistically analyzing and identifying differences in the Secure Sockets Layer (SSL) protocol. SSL uses a combination of public-key and symmetric key encryption. Each SSL connection always starts with the exchange of messages from the server and the client until a secure connection (handshake) is achieved. The handshake allows the server to prove its identity to the client using public-key encryption methods and then allows the client and server to work together to create a symmetric key to be used to quickly encrypt and decrypt the data exchanged between them. Optionally, the handshake also allows the client to prove his identity on the server. Since each Tor client generates self-signed SSL, using a random Algorithmically Generated Domain that changes every 3 minutes or so, a network traffic statistical analysis based on the specifics and characteristics of SSL can identify Tor sessions on a network combined with HTTPS traffic [8], [18], [23], [24].

In this paper, we propose a novel darknet traffic analysis and network management framework to real-time automating the malicious intent detection process, using a weight agnostic neural networks architecture. It is an effective and accurate computational intelligent forensics tool for network traffic analysis, the demystification of malware traffic, and encrypted traffic identification in real-time. Based on Weight Agnostic Neural Networks (WANNs) methodology, we propose an automated searching neural-net architecture strategy that can perform various tasks, such as identify zero-day attacks. By automating the malicious intent detection process from the darknet, the advanced proposed solution is reducing the skills and effort barrier that prevents many organizations from effectively protecting their most critical assets.

## 4. Literature Reviews

Dark Web is considered as a segment of the Deep Web (see Fig. 1), an intentionally hidden content that can be accessed with special software like Torbrowsers. Tor enables users to route their traffic through "users' computers" so in order that traffic is not traced back to the originating users and conceal their identity. To pass the data from one layer to another layer, Tor OR has created "relays" on computers that carry information through its tunnels all over the world. The encrypted information is placed between the relays. Tor traffic as a whole goes through three relays and then it is forwarded to the final destination [25].

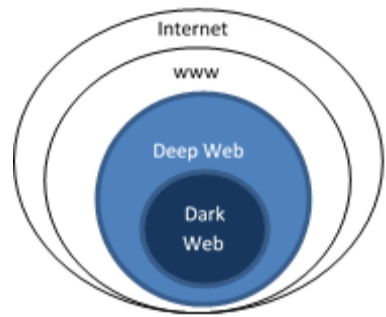

**Fig. 1.** The internet layers

There is an increasing interest in research related to the Dark web. A big part of the conducted literature review in cybersecurity is focused on anomaly-based network intrusion detection systems [9], [26]–[30]. In addition, there is research work dedicated in network traffic classification [31]–[33], whereas the Internet of Things (IoT) have recently attracted a significant amount of attention in machine learning and in network traffic analysis [34]–[37]. The yang et al. [21] introduce the current mainstream dark network communication system TOR and develop a visual dark web forum post association analysis system to graphically display the relationship between various forum messages and posters, and help law enforcement officers to explore deep levels. In addition this paper [38] designs a framework based on Hadoop is hidden threat intelligence. The framework uses a HBase-based distributed database to store and manage threat intelligence information and a web crawler is used to collect data through the anonymous TOR tool in order to identify the characteristics of key dark network criminal networks, which is the basis for the later dark network research.

A survey of different techniques and intrusion classification on KDD Cup 99 dataset was presented In [9] and an effective technique was suggested which categorised and identified intrusions in these datasets. In [26] a mapping of unlabeled trading data onto a set of two-dimensional grids and forming a set of bitmaps that identify anomalous and normal sessions. In the survey work of [27] a review is conducted in various intrusion detection models and methodologies for classification and data volume reduction. Most of these works used KDD-Cup 1999 dataset [28], or its successor NSL-KDD [6], which resolves some of the inherent issues of the first and has been widely adopted by the research community [29], [30]. However, the authors of [39] reported inefficiencies in most anomaly-based network intrusion detection systems employing supervised algorithms and suggested an unsupervised outlier detection scheme as a measure to overcome these inefficiencies. Other researchers suggested hybrid approaches for intrusion detection systems, with promising results, like for instance the authors of [40], who combine a Random-Forest classification technique and K-Means clustering algorithms and the authors of [41] who propose a combination of a deep autoencoder and an ensemble k-nearest neighbor graphs, based anomaly detector.

Concerning network traffic classification technologies, Bayesian Networks and Decision Trees algorithms were evaluated among others in [31] and found suitable for traffic flow classification at high speed. In [32] a systematic review of traffic classification approaches for machine learning was made, and a set of trends is derived from the analysis performed, whereas in [33] three major methods to classify different categories of Internet traffic are evaluated with their limitations and benefits. In [33] an hierarchical spatial-temporal features-based intrusion detection system (HAST-IDS) is proposed, which initially learns the low-level spatial network features of network traffic using deep

Convolutional Neural Networks (CNNs) and then learns high-level temporal features using long short-term memory networks. According to the authors the proposed scheme demonstrates low False Alarm Rate (FAR), high accuracy and detection rate. In [35] a fast and large-scale monitoring system is presented for monitoring the traffic on darknet consisting of two parts, pre-processing and classifier. In the pre-processing part, darknet packets are transformed into a feature vector consisting of 17 traffic features on darknet traffic. In classifier data fast online learning is actualized by training with traffic features of known Distributed Denial of Service (DDoS) attacks. The authors presented measurements results showing the proposed solution detects with high accuracy backscatter packets caused by DDoS attacks. It also adapts very quickly to new attacks.

On the other hand, recent research has demonstrated that the assumption that the data samples collected for training machine learning models are typically assumed to be independent and identically distributed can be problematic as it simplifies the manifold of structured data. This has motivated different research areas such as data poisoning, model improvement, and explanation of machine learning models [42]. The ability to explain in understandable terms, why a machine learning model makes a certain prediction is becoming immensely important, as it ensures trust and transparency in the decision process of the model. Shapley values provide accurate explanations, as they assign each feature an importance value for a particular prediction [43]. For example the authors of the [44] introduce a new metric, Top Similarity method, that measures the similitude of two given explanations, produced by Shapley values, in order to evaluate the Model-Agnostic Interpretability. Also the Florin [45]proposes a destructive method for optimizing the topology of neural networks based on the Shapley value, a game theoretic solution concept which estimates the contribution of each network element to the overall performance. More network elements can be simultaneously pruned, which can lead to shorter execution times and better results. An evolutionary hill climbing procedure is used to fine-tune the network after each simplification.

## 5. Methodology and Dataset

In recent years it has been shown that advanced Machine Learning Algorithms, such as neural networks, has the potential to be successfully applied in many areas of industry and the production process. Their success is based on the thorough processing of data that records the behavior of a system. By detecting patterns in the collected data, valuable information can be gleaned, future predictions can be made that automate a set of processes and provide serious impetus to the modern industry for value creation.

For example, multilayer neural networks, which are considered to be the easiest learning architectures, contain several linear layers that are laid out next to each other. Each of them takes an input from the previous level, multiplies it by some weights, adds a vector of bias to them, and passes the total vector through an activation function to produce the output of the level. This promotion process continues until the classification process is completed receiving the result from the final level. The final output is compared to the actual sorting values, where the sorting error is calculated using an appropriate loss function. To reduce the loss, the weights for all levels are updated one by one, using some Stochastic Gradient Descent.

Nevertheless, their application to realistic problems remains a very complex and specialized case. This is because data scientists, based on their hypotheses and experience, coordinate their numerous parameters, correlating them with the specific problems they intend to solve, utilizing the available training data sets. This is a long, tedious and costly task.

### 5.1 MetaLearning

MetaLearning is a recent holistic approach, which automates and solves the problem of specialized use of machine learning algorithms. It aims at the use of automatic machine learning to learn the most appropriate algorithms and hyperparameters that optimally solve a machine learning problem [46].

In particular, machine learning can be seen as a search problem, approaching an unknown underlying mapping function between input and output data. Design options, such as algorithms, model parameters (weights), hyper-parametric characteristics, and their variability, limit or expand the scope of possible mapping functions, i.e. search space.

It should be emphasized that the parameters of the model (weights) are implemented based on the data during the training, while the hyperparameters are determined by the developer before the training phase begins. They are not drawn from the data during training and this is the reason why they are usually stable. For example, the learning rate in a neural network is a hyperparameter. If this rate is too high, the network may exceed the local minimum, whereas if the learning rate is too low, the training may take a long time, as the steps taken during the descent are very small and also possible to be trapped in local minimums.

This logic is similar to various machine learning algorithms. What separates and distinguishes the algorithms from each other are their individual structural elements (e.g. in neural networks the structural elements are layers, in random forests they are the decision trees, etc.).

MetaLearning allows the training and comparison of one or more learning algorithms with different configurations of hyperparameters, data sets, optimization techniques, algorithm combinations, etc., in order to find the most appropriate learning model or set for a specific problem.

It is a subfield of machine learning where advanced learning algorithms are applied to the data and metadata of a given problem so that the models are learning to learn from previous learning processes or from previous classification tasks they have completed. It is an advanced form of learning where computer models, usually consisting of multiple subtraction levels, can improve the ability to learn by learning some or all of their own building blocks through the experience they gain in handling a large number of tasks. Their building blocks can be optimizers, loss functions, initializations, learning rates, updated functions, architectures, etc.

In general, for real physical problem modeling situations, input patterns with and without tags come from the same boundary distribution or follow some common cluster structure. Thus the data which has been already classified can contribute to the learning process, while respectively from the unclassified data can be extracted useful information for exploring the data structure of the general set, which can be combined with knowledge from previous learning processes or from previous classification tasks that have been processed [47].

Based on the above assumption, meta-learning techniques can discover the structure between data by allowing new tasks to be quickly learned using different types of metadata, such as the properties of the learning problem, the properties of the algorithm used (eg. performance measures), or patterns derived from data from a previous problem. In other words, they use cognitive information from unknown examples sampled from the distribution followed by the examples in the real world, in order to enhance the result of the learning process. In this way, it is possible to learn, select, change or combine different learning algorithms to effectively solve a given problem.

A meta-learning system should combine the following three requirements [47]–[50]:

1. The system must include a learning subsystem.
2. Experience has to be gained by utilizing the knowledge extracted from metadata related to the data set under process or from previous learning tasks that have been completed in similar or different fields.
3. Learning bias must be chosen dynamically.

Taking a holistic approach, a reliable meta-learning model should be trained in a variety of learning tasks and optimized for better performance in generalizing tasks, including potentially unknown cases. Each task is associated with a set of data D, containing attribute vectors and class tags on a supervised learning problem. The optimal parameters of the model are:

$$\theta^* = arg_{\theta}^{min} E_{D \sim P(D)}[L_{\theta}(D)] \quad (1)$$

which looks similar to a normal learning process, but a data set is considered a sample of data.

Data set D is often divided into two parts, a training set S and a set of B predictions for testing and testing,

$$D = \langle S, B \rangle \quad (2)$$

D datasets contain pairs of vectors and tags so that:

$$D = \{(x_i, y_i)\} \quad (3)$$

Each tag belongs to a known set of L tags.

We assume a classifier $f_{\theta}$. The parameter θ derives a probability of a data point belonging to the class y given by the attribute vector $x$, $P_{\theta}(y|x)$. Optimal parameters should maximize the likelihood of detecting true tags in multiple $B \subset D$ training batches:

$$\theta^* = argmax_{\theta} E_{(x,y) \in D}[P_{\theta}(y|x)] \quad (4)$$

$$\theta^* = argmax_{\theta} E_{B \subset D} \left[ \sum_{(x,y) \in B} P_{\theta}(y|x) \right] \quad (5)$$

The goal is to reduce the prediction error in data samples with unknown tags given that there is a small set of support for fast learning that works as fine-tuning.

It could be said that fast learning is a trick in which a fake data set is created that contains a small subset of tags (to avoid exposing all the tags in the model) and various modifications are made to the optimization process in order to achieve the fast learning. A brief step-by-step description of the whole process is presented below:

1. Creation of a subset of $L_s \subset L$ tags.
2. Creation of an $S^L \subset D$ training subset and a $B^L \subset D$ prediction set. Both of these subsets include labeled data belonging to the subset $L_s$, y∈$L_s$,∀(x,y)∈ $S^L$, $B^L$.
3. The optimization process uses the $B^L$ subset to calculate the error and update the model parameters via error backpropagation, in the same way, that it is used in a simple supervised learning model.

In this way, it can be considered that each sample pair $(S^L, B^L)$ is also a data point. Thus the model is trained so that it can generalize to new, unknown data sets.

A modification of the supervised learning model is the following function, to which the symbols of the Meta-Learning process have been added:

$$\theta^* = argmax_{\theta} E_{L_s \subset L} \left[ E_{S^L \subset D, B^L \subset D} \left[ \sum_{(x,y) \in B^L} P_{\theta}(x, y, S^L) \right] \right] (6)$$

It should be noted that retrospective neural networks with only internal memory, such as Long Short-Term Memory (LSTM), are not considered meta-learning techniques. On the contrary, the most appropriate meta-learning architecture for neural networks is Neural Architecture Search (NAS) [51].

### 5.2 Neural Architecture Search

It is an automated learning technique for automating the design of artificial neural networks, the most widely used field in the field of machine learning. NAS has been used to design networks that are equivalent to or superior to hand-drawn architectures.

NAS methods can be categorized according to the search space, search strategy, and performance estimation strategy used [51]–[53]:

1. The search area determines the types of neural networks that can be designed and optimized in order to find the optimal type of neural network that can solve the given problem e.g. Forward Neural Network (FFNN), Recurrent Neural Network (RNN), etc.

2. The search strategy determines the approach used to explore the search space, i.e. the structure of the architectural design in an internal search field of hyperparameters (levels, weights, learning rate, etc.).

3. Performance appraisal strategy evaluates the performance of a potential neural network by designing it without constructing and training it.

In many NAS methods, both micro and macro structures are searched hierarchically, allowing the exploration of different levels of standard architectures. The three NAS strategies and hierarchical search methods are shown in the Fig. 2.

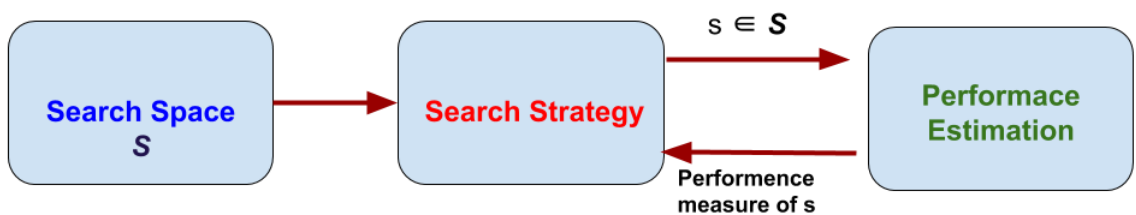

**Fig. 2** The three Neural Architecture Search strategies

In hierarchical search, the first level consists of the set of primitive functions, the second level of different patterns that connect primitive functions through a directed acyclic graph, and the third level of patterns that encode how the second level patterns are connected, and so on.

NAS is closely related to hyper-parameter optimization and is a subfield of automated machine learning designed to follow best practices for reducing program load, providing stable and simple environments, minimizing the number of actions required for use, providing a clear methodology for discovering knowledge in unfamiliar environments [54].

Specifically, given a neural architecture search space F, where the input data D is divided into $D_{train}$ and $D_{val}$ and the cost function Cost ($\cdot$) (eg accuracy, mean squared error, etc), the goal is to find an optimal neural network $f^* \in F$, which can achieve the lowest cost in the data set D.

Finding the optimal neural network $f^*$ is equivalent to:

$$f^* = argmin_{f \in F} \, Cost(f(\theta^*), D_{val}) \quad (7)$$

$$\theta^* = argmin_\theta \, L(f(\theta), D_{train}) \quad (8)$$

$\theta^*$ is the learning parameter of the network.

A simplified NAS procedure is described in the following code:

```
def search(self):
    # for the number of controller epochs
    for controller_epoch in range(controller_sampling_epochs):
        # sample a set number of architecture sequences
        sequences=sample_architecture_sequences(controller_model, samples_per_controller_epoch)
        # predict their accuracies using a hybrid controller
        pred_accuracies = get_predicted_accuracies(controller_model, sequences)
        # for each of these sequences
        for i, sequence in enumerate(sequences):
            # create and compile the model corresponding to the sequence
            model = create_architecture(sequence)
```

```
                    # train said model
                    history = train_architecture(model)
                    # log the training metrics
                    append_model_metrics(sequence, history, pred_accuracies[i])
            # use this data to train the controller
            xc, yc, val_acc_target = prepare_controller_data(sequences)
            train_controller(controller_model, xc, yc, val_acc_target)
```

Given that there can be theoretically infinite possibilities for compiling different hyperparameters in the search space, it is legitimate to create the appropriate constraints that will lead to the optimization of the problem. For example, the number of neurons in each hidden level can only be a positive integer. Also, restrictions can be placed on the use of activation functions, as they serve different purposes and functions e.g. it is indicated the use of a sigmoid activation function if the classification problem is binary.

In particular, the following should be considered for the precise design of the NAS strategy [51], [53]:

1. How many neurons should it contain in each hidden level. There are countless options and it must be found which configuration will give the best categorization accuracy.

2. Which activation function works best for each hidden level. There are many different activation functions, so finding the optimal one for a particular data set requires experimentation which should be automated.

3. Addition of a drop layer. It should be investigated whether it helps or damages the performance of the architecture.

4. What is the function of the final level. It should be investigated whether this is a multi-order problem or a binary classification problem. This determines the number of nodes in the final level, as well as the loss function that can be used in architectural training.

5. The dimension of data. If two-dimensional or three-dimensional input is needed and if there is a need for normalization before linear levels are added.

6. How many hidden layers will eventually be used. They rarely need more than two hidden levels in a neural network for optimal performance when the problem lies in multidimensional data analysis areas (photos, video, audio, etc.).

The design of the NAS strategy has as its primary objective the definition of a neural network architecture that adapts to the nature of the data set under consideration and to the precise coordination of ideal hyperparameters that can lead to a model with high accuracy and generalizability to data outside of training and testing sets.

Typical hyperparameters that can be optimized and need to be tuned include optimization algorithms (SGD, Adam, etc.), learning rate programming, regularization, etc [51]–[53]. Essentially, they enable the creation of the best learning techniques with high-performance success, with very little effort and minimal know-how.

The way the NAS strategy works can be bidded with techniques based on the ways in which nature works, and in particular on finding proportions between techniques where instinct, as a sexual characteristic, prevails over education. For example, some species in biology have predatory behaviors from the moment of their birth, which allows them to perform complex motion and sensory tasks without learning, which in many cases are completely satisfactory for the survival of the species. In contrast, in the training of artificial neurons to perform a task, an architecture that is considered suitable for modeling the task is usually chosen, and the search focuses mainly on finding the weight parameters using a learning algorithm. Inspired by social behaviors that evolved in nature, neural networks can be developed with architectures that are naturally capable of performing a given task even when their weight parameters are randomized, so they can perform well without training, while their performance can be further maximized through training.

### 5.3 Proposed Method

A Weight Agnostic Neural Networks (WANN) methodology [55] was used in this paper. It is an evolving strategy in neural network development techniques that can perform a specialized task regardless of the weights of the connections in the building blocks of the neural network, which equates to a lack of training [56]. The logic of using WANN is a basic investigation in the search for architectural neural networks with specific biases that can potentially categorize the given problem, even when using random weights. By exploring such architectures, it is possible to explore factors that can perform well in their interaction environment without the need for training, which is a digital security system that can create robust self-identifying systems capable of identifying zero-day attacks.

The investigation of the WANN architectural structures used in this research, fully exploits the theoretical approach of Inductive Bias for the production of biased results, due to the assumptions/choices made either for the representation of the case space or for the definition of the search engine in the field of assumptions. The unrestricted space of cases is potentially infinite. By choosing the type of knowledge, i.e. the representation, this space is limited and the first level of bias is introduced. Even so, the search space is probably still too large to perform a full search. Thus the second level of bias is introduced, that of the search algorithm e.g. the algorithm used does not perform a complete search in the area of possible solutions but approaches it heuristically or probabilistically. However, without these options, a learning algorithm would not be better than a random selection algorithm.

Also, the training data used is finite, so they do not accurately reflect the reality, as the selection process and the assumption that this data will have the same distribution as in all cases introduces another level of bias. So by reducing the inductive learning hypothesis, it is implied that the model produced by a limited set of training data will describe new, unknown cases just as well.

In conclusion, each learning algorithm has a specific bias in its elements such as representation, algorithm, or data and this is a fundamental, necessary feature, which is taken seriously into the investigation of appropriate architectural neural networks.

Another very important process that will allow the identification of appropriate architectures to solve the given problem, concerns the interpretability of the methods used. Global interpretability offers a holistic picture of the model. It is about understanding how the model makes decisions, what its most important features are and what interactions take place between the features. A full understanding is very difficult to achieve in practice therefore, the explanations at the universal level concern a general global representation of the model, which is not detailed and accurate. Similarly, when the explanations focus on a small area of data, then there is local interpretability, where a single example of the data set is analyzed and it is explained why the model made a specific decision about it. In small areas of data, the prediction may depend only linearly or monotonously on certain features, instead of having a more complex dependence on them.

Shapley values [42], [43]are a very effective way of generating explanations from Cooperative / Calitional Game Theory. The payoff/gain of the players of a cooperative game is given by a real function that gives values to sets of players. The Shapley value is the only payoff function that satisfies 4 key properties:

1. Anonymity: The axiom of anonymity states that the order of the players does not affect the amount allocated to them by the Shapley value. In other words, the above relation says that if we shift the players, then the Shapley value of the new game does not result from the corresponding shift of the Shapley value coordinates of the original game. A consequence of the axiom of anonymity is the axiom of symmetry, which states that the Shapley values of two symmetric players 1 and 3 are equal.

2. Efficiency: The axiom of efficiency determines that the distribution of social wealth according to the Shapley value is collectively rational.

3. Zero-Useless: This axiom determines that if a player has zero contribution to social wealth, then his Shapley value is 0.

4. Additionality: This axiom specifies that the Shapley value of the sum game is the sum of the Shapley values.

The connection of Shapley values to the problem of explaining WANN architectural structures is done in the following way. We consider the problem of WANN architectural structures as a cooperative game, whose players are the characteristics of the data set, the profit function is the neural network model under consideration and the model predicts the corresponding profits. In this context, the Shapley values show the contribution of each feature and therefore the explanation of why the model made a specific decision.

In conclusion, the Shapley value of characteristic $i$ of a neural network model $f$ is given by the following equation [57]:

$$\varphi_i = \sum_{S \in F \setminus \{i\}} \frac{|S|! \, (M - |S| - 1)!}{M!} \left[ f_{S \cup \{i\}} (x_{S \cup \{i\}}) - f_S(x_S) \right]$$

where F is the set of attributes, S is a subset of $F$ and $M = |F|$ is the sum of the set $F$. This relation measures the weight of each attribute by calculating its contribution when it is present in the prediction and then subtracts it when it is absent,

More specifically::

1. $f_{S \cup \{i\}}(x_{S \cup \{i\}})$: is the output when the $i^{\infty}$ attribute is present.
2. $f_S(x_S)$: is the output when the $i^{\infty}$ attribute is not present.
3. $\sum_{S \in F \setminus \{i\}} \frac{|S|!(M-|S|-1)!}{M!}$: is the weighted mean of all possible subsets $S$ in $F$.

The SHapley Additive exPlanations (SHAP) [58] method explains model decisions using Shapley values [59]. An innovation of SHAP is that it works as a linear model and more specifically as a method of additional contribution of features.

Intuitively with the SHAP approach, an explanation is a local linear approach to model behavior. In particular, while the model can be very complex as an integrated entity, it is easy to approach a specific presence or absence of a variable. For this reason, the degree of linear correlation of the independent and dependent variables of the set with dispersion $\sigma_X^2$ and $\sigma_Y^2$, respectively, and the covariance $\sigma_{XY} = Cov(X, Y) = E(X, Y) - E(X)E(Y)$, which is measured by calculating the *Pearson R* correlation table, and is defined as follow:

$$R = \frac{\sigma_{XY}}{\sigma_X \sigma_Y}$$

However, given the inability of the above method to detect nonlinear correlations such as sinus wave, quadratic curve, etc. or to explore the relationships between the key variables, the Predictive Power Score (PPS) technique [60] was selected and used in this study for the predictive relationships between available data. PPS, unlike the correlation matrix, can work with non-linear relationships, with categorical data, but also with asymmetric relationships, explaining that variable A informs variable B more than variable B informs variable A. Technically, scoring is a measurement in the interval [0, 1] of the success of a model in predicting a variable target with the help of an off-sample variable prediction, which practically means that this method can increase the efficiency of finding hidden patterns in the data and the selection of appropriate forecast variables.

The use of the PPS method also focuses on the fact that a local explanation must be obtained of the models that are initially capable of operating without training and after being reinforced at a later time with training. However, the sensitivity of the SHAP method to explain the models in their hyper-parameter values, as well as the general inability to deal with the high data dimension, requires the implementation of feature selection before the application of the technique. In particular, the complexity of the problem in combination with the large number of explanations that must be given for the predictions of the model, is significantly more difficult, as the distinction between relevant and irrelevant features, as well as the distances between data points, can not be fully captured.

Taking this observation seriously, feature selection was performed to optimally select a subset of existing features without transformation, in order to retain the most important of them, in order to reduce their number and at the same time retaining as much useful information as possible. This step is crucial because if features with low resolution are selected, the resulting learning system will not perform satisfactorily, while if features that provide useful information are selected, the system will be simple and efficient. In general, the goal is to select those characteristics that lead to long distances between classes and small variations between the same class.

The process of feature selection was done with the PPS technique, where for the calculation of PPS in numerical variables the metric of Mean Absolute Error (MAE) was used which is the measurement of the error between the estimation or prediction in relation to the observed values and is calculated below [61]:

$$MAE = \frac{1}{n}\sum_{i=1}^{n} |f_i - y_i| = \frac{1}{n}\sum_{i=1}^{n} |e_i|$$

where $f_i$ is the estimated value and $y_i$ is the true value. The average of the absolute value of the quotient of these values is defined as the absolute error of their relation $|e_i| = |f_i - y_i|$.

Rescue and Precision F-Score (Harmonic Mean) was used for the categorical variables, respectively, implying that the higher the F-Score, the higher the two metrics respectively. The calculation is accomplished from the following relation [61]:

$$F_{Score} = \frac{2 \times recall \times precision}{recall + precision} = \frac{2TruePositives}{2TruePositives + FalsePositives + FalseNegatives}$$

In conclusion, to create a cybersecurity environment with fully automated solutions capable of recognizing content from the Darknet, a NAS development strategy was implemented, based on the WANN technique which is reinforced with explanations with Strapley values, having first preceded feature selection process with the PPS method

### 5.4 Dataset

Darknet, as an overlay network, is only accessed with specific software, configuration, or licenses, often using non-standard communication protocols and ports. Its address space is not accessible for interaction with familiar web browsers and any communication with Darknet is considered skeptical due to the passive nature of the network in managing incoming packets.

The classification of Darknet traffic is very important for the categorization of real-time applications, while the analysis of this traffic helps in the timely monitoring of malware before the attack, but also in the detection of malicious activities after the outbreak.

The data set used in this study is based on CICDarknet2020, which includes Darknet traffic as well as corresponding normal traffic from Audio-Stream, Browsing, Chat, Email, P2P, Transfer, Video-Stream, VOIP, Files, Session and Authentication, which are implemented or not over Tor and VPN infrastructures. The following Table 1 provides details of the final categories used and the applications that implement them [62].

**Table 1.** Darknet Network Traffic Details

| ID | Traffic Category | Applications Used |
|----|------------------|-------------------|
| 0 | Audio-Stream | Vimeo and Youtube |
| 1 | Audio-Stream | Crypto streaming platform |
| 2 | Browsing | Firefox and Chrome |
| 3 | Chat | ICQ, AIM, Skype, Facebook and Hangouts |
| 4 | Email | SMTPS, POP3S and IMAPS |
| 5 | P2P | uTorrent and Transmission (BitTorrent) |
| 6 | File Transfer | Skype, SFTP, FTPS using Filezilla and an external service |
| 7 | File Transfer | Crypto transferring platform |
| 8 | Video-Stream | Vimeo and Youtube |
| 9 | Video-Stream | Crypto streaming platform |

| 10 | VOIP | Facebook, Skype and Hangouts voice calls |

## 6. Experiments and Results

Initially and in order to have a level of comparison of the proposed methodology, the above dataset was used in the investigation of solutions through the most well-known machine learning methods, in the identification and categorization of network traffic in Tor, Non-Tor, VPN, and Non-VPN of the Table services. 1. The results of the categorization process are presented in detail for each algorithm, in the following Table 2

**Table 2.** Classification Performance Metrics

| Classifier | Accuracy | AUC | Recall | Precision | F1 | Kappa | MCC | TT (Sec) |
|---|---|---|---|---|---|---|---|---|
| *Extreme Gradient Boosting (XGB)* | 0.9012 | 0.9953 | 0.7500 | 0.9014 | 0.8990 | 0.8751 | 0.8756 | 441.61 |
| *CatBoost* | 0.8927 | 0.9942 | 0.7227 | 0.8936 | 0.8894 | 0.8642 | 0.8648 | 606.07 |
| *Decision Tree* | 0.8858 | 0.9477 | 0.7406 | 0.8845 | 0.8849 | 0.8561 | 0.8562 | 1.90 |
| *Random Forest* | 0.8846 | 0.9848 | 0.7245 | 0.8829 | 0.8835 | 0.8545 | 0.8545 | 19.83 |
| *Gradient Boosting* | 0.8801 | 0.9916 | 0.7106 | 0.8797 | 0.8764 | 0.8482 | 0.8488 | 645.38 |
| *Extra Trees* | 0.8775 | 0.9677 | 0.7201 | 0.8756 | 0.8762 | 0.8455 | 0.8455 | 11.61 |
| *K Neighbors* | 0.8504 | 0.9663 | 0.6748 | 0.8462 | 0.8466 | 0.8105 | 0.8108 | 7.45 |
| *Light Gradient Boosting Machine* | 0.7826 | 0.8986 | 0.5387 | 0.7911 | 0.7808 | 0.7247 | 0.7259 | 17.41 |
| *Ridge Classifier* | 0.6664 | 0.0000 | 0.3276 | 0.6672 | 0.6221 | 0.5659 | 0.5727 | 0.40 |
| *Linear Discriminant Analysis* | 0.6497 | 0.9136 | 0.4400 | 0.6439 | 0.6231 | 0.5535 | 0.5575 | 2.01 |
| *Quadratic Discriminant Analysis* | 0.3858 | 0.8710 | 0.4026 | 0.6325 | 0.4394 | 0.2936 | 0.3144 | 0.71 |
| *Logistic Regression* | 0.3174 | 0.6756 | 0.1226 | 0.3089 | 0.2753 | 0.1237 | 0.1433 | 126.46 |
| *Naïve Bayes* | 0.2974 | 0.6328 | 0.1281 | 0.2303 | 0.2278 | 0.0960 | 0.1120 | 0.13 |
| *SVM − Linear Kernel* | 0.1937 | 0.0000 | 0.1037 | 0.2419 | 0.1248 | 0.0379 | 0.0485 | 130.74 |
| *Ada Boost* | 0.1626 | 0.7142 | 0.1521 | 0.0501 | 0.0713 | 0.0788 | 0.1193 | 14.27 |

*\*AUC=Area Under the Curve, MCC=Matthews Correlation Coefficient, TT=Training Time*

Also, the Receiver Operating Characteristic (ROC) curves, the Confusing Matrix, and the Class prediction error diagram of the XGBoost method, which achieved the highest success results (Accuracy 90%), are presented in Figs 3-5.

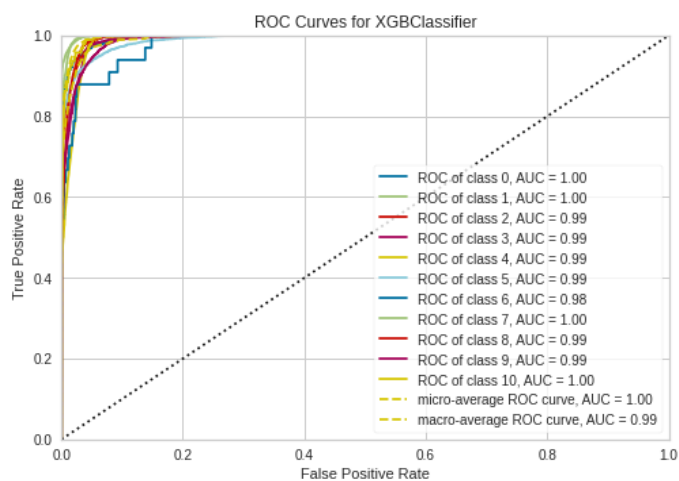

**Fig. 3.** Receiver Operating Characteristic curves of the XGBoost classifier

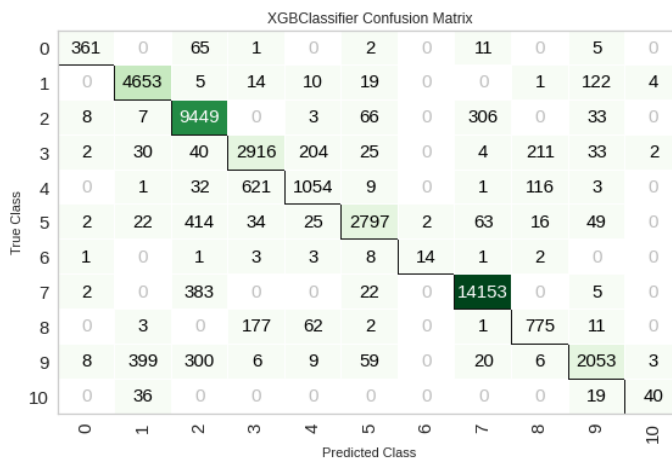

**Fig. 4.** Confusing Matrix of the XGBoost classifier

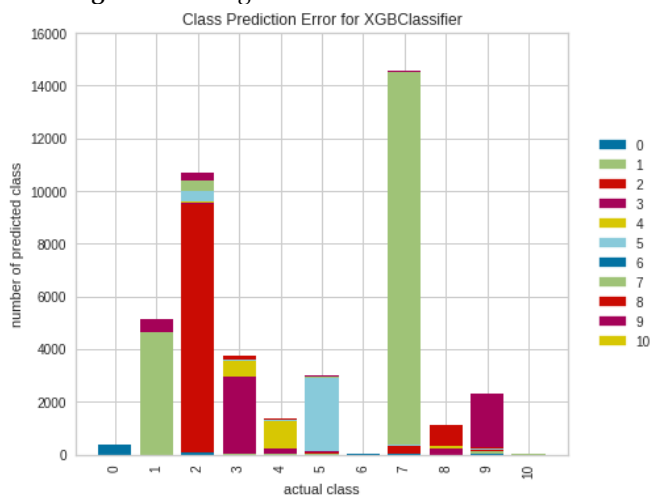

**Fig. 5.** Class prediction error of the XGBoost classifier

As evidenced by the above results, the problem of analysis and categorization of network traffic is a rather complex problem, which although can not be considered multidimensional, the distinction between the characteristics of the data set, as well as the distances between points are considered very dense, which makes it significantly difficult to implement distinct levels of decision and therefore categorization with very high results.

The automated creation of network architectures that encode search solutions through NAS can produce architectures that, once trained, go beyond human-designed ones. With the architecture in question and specifically with the methodology based on the Autokeras NAS library [59], which is designed to provide stable and simple interface environments, minimizing the number of user actions, the following Table 3 architecture was implemented, which is shown in the figure. 4, with the results of Table 4.

**Table 3.** AutoKeras Model

| Layer (type) | Output Shape | Parameters |
|---|---|---|
| input_1 (InputLayer) | [(None, 61)] | 0 |
| multi_category_encoding | (Mul (None, 61) | 0 |
| normalization | (Normalization (None, 61) | 123 |
| dense (Dense) | (None, 512) | 31744 |
| re_lu (ReLU) | (None, 512) | 0 |
| dense_1 (Dense) | (None, 128) | 65664 |
| re_lu_1 (ReLU) | (None, 128) | 0 |
| dense_2 (Dense) | (None, 11) | 1419 |
| classification_head_1 | (Softm (None, 11) | 0 |
| | **Total params:** | **98,950** |
| | **Trainable params:** | **98,827** |
| | **Non-trainable params:** | **123** |

**Fig 4.** Depiction of the AutoKeras Model

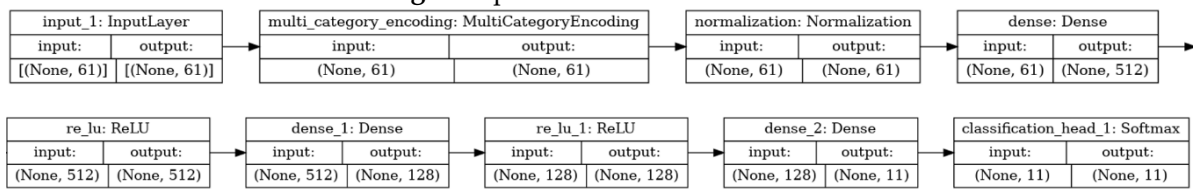

**Table 4.** Classification Performance Metrics of the AutoKeras Model

| Classifier | Accuracy | AUC | Recall | Precision | F1 | Kappa | MCC | TT (Sec) |
|---|---|---|---|---|---|---|---|---|
| AutoKeras Model | 0.9268 | 0.9976 | 0.7915 | 0.9280 | 0.9105 | 0.8972 | 0.8977 | 917.23 |

The networks in question created by NAS, although much slower and more complex (Trainable parameters: 98,827), prove to be excellent after training, as evidenced by the results of the table above.

In no case, however, can we assume that they are able to solve the given problem without training their weights. To produce architectures that satisfactorily encode solutions, the importance of weights must be minimized. Instead of judging networks by their performance with optimal weight values, they should be evaluated by their performance when their weight values come from a random distribution. Replacing weight training with weight sampling ensures that performance is only a product of network topology and not training.

The architecture chosen to solve the given problem by a random selection of weights involves a recurring neural network with input, some sparsely connected hidden layers of reservoirs in which the choice of architecture is based on the NAS strategy, and a simple linear readout output. The connection weights in each reservoir, as well as the input weights, are random and scaled in such a way as to ensure the Echo State Property property which is defined as a state so that the reservoir is an "echo" of the entire input history and which is partly determined by its architecture.

The discrete levels are only those of the $u(n)$ input and the $y(n)$, output, while the hidden levels (reservoirs) are grouped so that the neurons are connected to each other by a percentage that determines how sparsity the network is. The synaptic compounds that unite the levels with each other are characterized by a value that determines the weights. Each input neuron is connected via $Win_{ij}$ weights (i-input neuron, j-neuron to the reservoir) with weights that although normal, are determined randomly prior to training and their values are final as they do not change during training. Also, each neuron from the reservoir is connected via $W_{jk}$ weights (j-neuron in the reservoir, k-neuron in the reservoir, and $j\neq k$) to any other neuron in the reservoir. The weights of these neurons although normal are determined randomly before training and their values do not change. Finally, each neuron from the reservoir is connected via $Wout_{jm}$ weights (j-neuron in the reservoir, m-output neuron) to the output plane neurons. These weights that are in the readout layer, are the only ones that are trained in order to get their final values [64].

The network architecture is characterized by a stacked hierarchy of reservoirs, where at each time step $t$, the first repeating layer is fed by the external input $u(t)$, while each successive layer is fed by the output of the previous one in the stack [65]. Although their architectural organization allows for general flexibility in the size of each layer, for reasons of complexity we consider a hierarchical installation of reservoirs with repeating layers $N_L$, each of which contains the same number of $N_R$ units. In addition, we use $x^{(l)}(t) \in R^{N_R}$ to declare the state of the plane $l$ at time $t$. By omitting the bias conditions, the first level state transition function is defined as follows [66]:

$$x^{(1)}(t) = \left(1 - a^{(1)}\right)x^{(1)}(t-1) + a^{(1)} \, tanh \, tanh \, \left(W_{in}u(t) + \widehat{W}^{(1)}x^{(1)}(t-1)\right)$$

For any level greater than $l>1$ the equation is as follows:

$$x^{(l)}(t) = \left(1 - a^{(l)}\right)x^{(l)}(t-1) + a^{(l)} \, tanh \, tanh \, \left(W^l x^{l-1}(t) + \widehat{W}^{(l)}x^{(l)}(t-1)\right)$$

where $W_{in} \in R^{N_R \times N_U}$ is the input weight table, $\widehat{W}^{(l)} \in R^{N_R \times N_R}$ is the recurrent weight table for level l, $W^{(l)} \in R^{N_R \times N_R}$ is the table relative to the connection weights between the levels from level $l$-1 to level $l$, $a^{(l)}$ is the leaky parameter at level $l$ and $tanh$ represents the elementary application of the tangent [67]–[70].

Random weights improve the generalization properties of the solution of a linear system because they produce almost rectangular (weakly correlated) features. Since the output of a linear system is always correlated with the input data, if the range of solution weights is limited, rectangular inputs provide a wider range of solutions than those supported by weights. Also, small weight fluctuations allow the system to become more stable and noise resistant, as input errors will not be amplified at the output of a linear system with little correlation between input and output weights. Thus the random classification of weights that produces weakly correlated characteristics at the latent level allows achieving a satisfactory solution and a good generalization performance.

Essentially for a random forward propagation architecture with a hidden plane and random representation of hidden plane neurons, the input data is mapped to a random L-dimensional space with a distinct set of training $N$ where $(x_i, t_i), i \in [\![1, N]\!]$ με $x_i \in R^d$ και $t_i \in R^c$. The network output is represented as follows[68], [71]:

$$f_L(x) = \sum_{i=1}^{L} \beta_i h_i(x) = h(x)\beta \quad i \in [\![1, N]\!]$$

where $\beta = [\beta_1, \dots, \beta_L]^T$ is the output of the weight table between the hidden nodes and the output nodes, $h(x) = [g_1(x), \dots, g_L(x)]$ are the outputs of the hidden nodes (random hidden attributes) for input $x$, and $g_1(x)$ is the exit of i hidden node. The basis of an N set of training $\{(x_i, t_i)\}_{i=1}^{N}$, can solve the learning problem Hβ = T, where $T = [t_1, \dots, t_N]^T$ the target labels and the output table of the hidden level $H$ as below:

$$H(\omega_j, b_j, x_i) = [g(\omega_1 x_1 + b_1) \cdots g(\omega_l x_1 + b_l) \vdots \ddots \vdots g(\omega_1 x_N + b_1) \cdots g(\omega_l x_N + b_l)]_{N \times l}$$

Prior to training, the input weight table $\omega$ and the bias vectors $b$ are randomly generated in the interval [−1, 1], with $\omega_j = [\omega_{j1}, \omega_{j2}, \dots, \omega_{jm}]^T$ and $\beta_j = [\beta_{j1}, \beta_{j2}, \dots, \beta_{jm}]^T$. The output level table of the hidden level $H$ is calculated from the activation function and the use of the training data based on the following function:

$$H = g(\omega x + b)$$

Respectively the output weights $\beta$ can be calculated from the relation:

$$\beta = \left(\frac{I}{C} + H^T H\right)^{-1} H^T X$$

where $H = [h_1, \dots, h_N]$ are the outputs of the hidden level and is the input data. $\beta$ can be calculated from the generalized inverse Moore-Penrose table:

$$\beta = H^+ T$$

where $H^+$ is the generalized inverse Moore-Penrose table for table $H$.

In this case, the proposed standardization offers the possibility of managing multiple intermediate representations, as the hierarchical organization of random reservoirs architecture in successive layers, naturally reflects the structure of the dynamics of the developed system. This scaling allows the progressive classification and exploration of input data interfaces across the levels of the hierarchical architecture, even if all levels share the same weight values. Furthermore, the multilevel architecture represents a transitional state of how the internal representations of the input signals are determined, which guarantees high performance even for problems that require long internal memory intervals. It also has higher performance in cases where short-term network memory capabilities are required than the corresponding architectures, which would have to work with the same total number of iterative or retrospective units in order to achieve corresponding results. In addition, in terms of computational efficiency, the multilevel construction of reservoirs in the design of a neural system also results in a reduction in the number of non-zero repetitive connections, typically associated with other types of retrospective architectures. This implies low complexity and time savings required to perform specialized tasks.

These conclusions are reflected in Table 5 below, which shows the very high categorization results (Accuracy 94%) in combination with the very short processing time (290sec which is 66% faster than the corresponding AutoKeras Model) :

**Table 5.** Classification Performance Metrics of the Reservoir Model

| Classifier | Accuracy | AUC | Recall | Precision | F1 | Kappa | MCC | TT (Sec) |
|---|---|---|---|---|---|---|---|---|
| Reservoir Model (13-11-09) | 0.9451 | 0.9988 | 0.8122 | 0.9317 | 0.9242 | 0.9108 | 0.9094 | 290.08 |

But even in this case, there was weight training and therefore the result of the process is a product of training. Unlike WANN, weight training is avoided. Focusing exclusively on exploring solutions in the field of neural network topologies, using random common weights for each network level, and recording the cumulative result during the test, reservoir architecture was used, without weight training. To identify the number of resulting solutions, the process is assisted with explanations using the Shapley value methods, after first selecting features with the PPS method. The resulting network population is then ranked according to their performance and complexity so that the highest-ranking networks are selected to form a new solution population. The process is repeated until the best architecture is found. The architectures are modified either by inserting a node by separating an existing connection, by adding a connection by connecting two previously unconnected nodes, and by changing the activation function which reassigns activation functions.

Initially, the predictive power of the problem variables was analyzed to identify the variables with the highest PPS, in order to identify the most important ones that can solve the problem, simplifying the process and at the same time without reducing the effectiveness of the method. From the total of variables, 19 were selected with a significant score greater than 0.3, while the rest had a predictive capacity of less than 0.1. A summary of the 19-variable PPS capture table is presented in Table 6.

**Table 6.** Predictive Power Score

| Idle_Max | Idle_Mean | Idle_Min | Packet_Length_Max |
|---|---|---|---|
| 0.471 | 0.444 | 0.430 | 0.399 |
| **Packet_Length_Mean** | **Average_Packet_Size** | **Flow_IAT_Max** | **Fwd_IAT_Max** |
| 0.383 | 0.379 | 0.372 | 0.363 |
| **Bwd_Packet_Length_Max** | **Fwd_Packet_Length_Max** | **Total_Length_of_Bwd_Packet** | **Bwd_Packet_Length_Mean** |
| 0.349 | 0.345 | 0.340 | 0.338 |
| **Bwd_Segment_Size_Avg** | **Total_Length_of_Fwd_Packet** | **Packet_Length_Std** | **Packet_Length_Variance** |
| 0.338 | 0.338 | 0.330 | 0.330 |
| **Fwd_Header_Length** | **Subflow_Bwd_Bytes** | **Fwd_Packet_Length_Mean** | |
| 0.328 | 0.320 | 0.313 | |

Extensive research was then conducted on evaluating the values of the variables, how they contribute to the prediction, and explaining each decision of the implemented models, using the Shapley values. Figure 4 shows the classification of the values of the variables used in the bar plot.

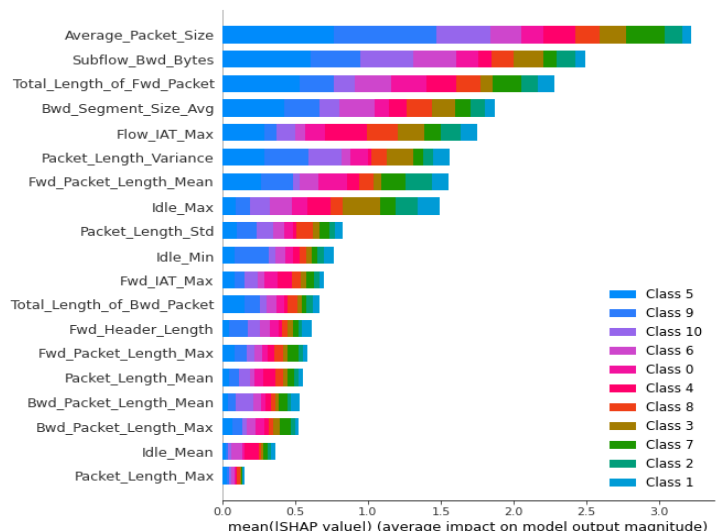

**Fig. 6.** Shap value impact on model output (bar plot)

Respectively in the following image is presented the summary beeswarm plot, which is the best way to capture the relative effect of all the features in the whole data set. Characteristics are classified based on the sum of Shapley values in all samples in the set. The most important features of the model are shown from top to bottom. Each attribute consists of dots, which symbolize each attribute of the package, while the color of the dot symbolizes the value of the attribute (blue corresponds to a low value, while red to a high value). The position of the dot on the horizontal axis depends on its Shapley value.

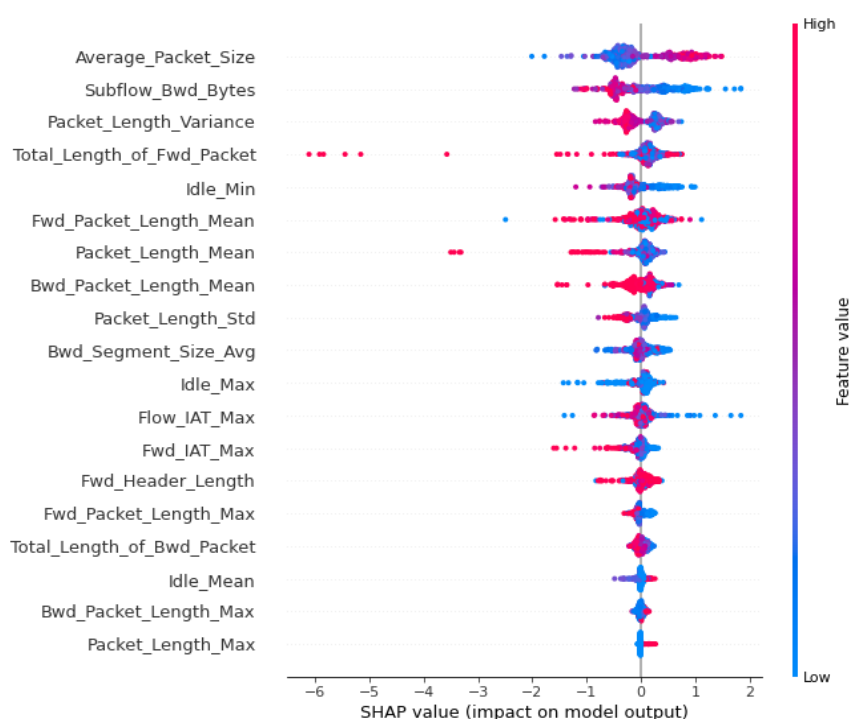

**Fig. 7.** Shap value impact on model output

We see that the Average_Packet_Size attribute is the most influential for the model predictions. Also for its high values (red dots), the Shapley value is also high so it has a great positive effect, i.e. it increases the probability that the package under consideration comes from Darknet. On the contrary, for its low values (blue dots), the Shapley value is low so it has a negative effect on the forecast, i.e. it increases the probability that the package under consideration does not come from Darknet.

In the image below a sample, selection is used from the data set to represent the typical attribute values and then 10 samples are used to estimate the Shapley values for a given prediction. This task requires 10 x 1 = 10 evaluations of the model.

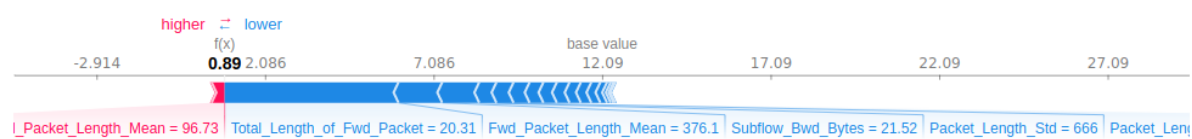

**Fig. 8.** Explain a single prediction (10 evaluations)

In this diagram a local explanation is presented, where the base value refers to the average value of the model forecasts, i.e. in this case the model predicts with a probability of 12% the data package analyzed to come from Darknet. For this package, the forecast value is 89%, so the Shapley values show the change from the average forecast to the specific forecast. The red arrows push the prediction to the right, that is, they help to increase the probability that this package comes from Darknet, while the blue arrows push to the left, helping to reduce the probability that it comes from Darknet. The length of each arrow symbolizes the magnitude of the effect on the prediction. In this example, we see that the Fwd_Packet_Length_Mean attribute helps to increase the likelihood that the package will come from Darknet (Shapley value 96.73), while the Total_Length_of_Fwd_Packet (Shapley value 20.31) and Fwd_Packet_Length_Mean (Shapley value 376.1), decrease this likelihood, etc.

Staplery values also have universal explanation capabilities, summing the values of a set of samples. In the image below you use a selection of 100 samples from the data set to represent the standard attribute values and then 500 samples are used to estimate the Shapley values for a given prediction. This task requires 500 x 100 = 50,000 model evaluations.

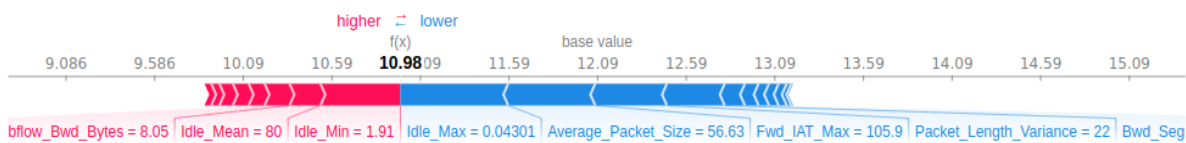

**Fig. 9.** Explain a single prediction (50,000 evaluations)

Respectively in figure 10 there is the diagram of the above process of 50,000 validations, but with an explanation of multiple predictions and their fixation in relation to their similarity

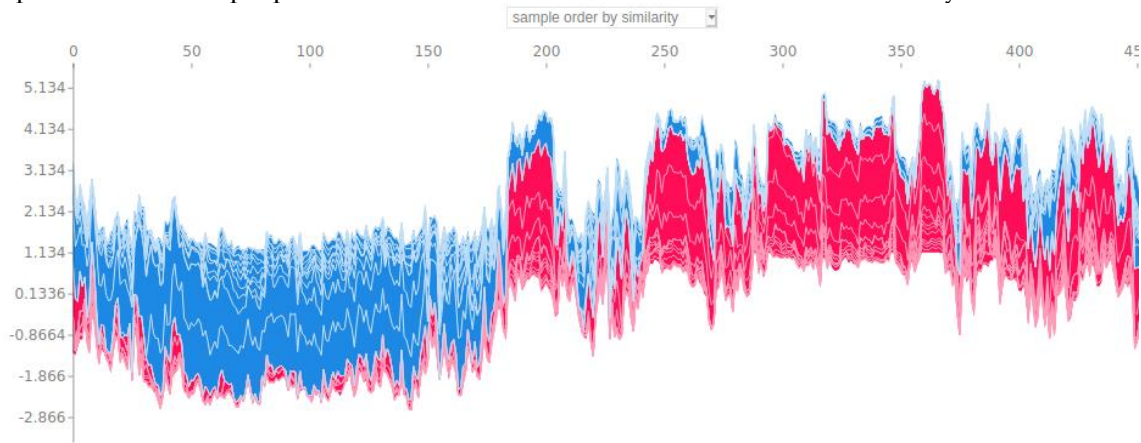

**Fig. 10.** Explain many predictions by similarity

Respectively in Fig.11 the same procedure is captured based on the output values of the model.

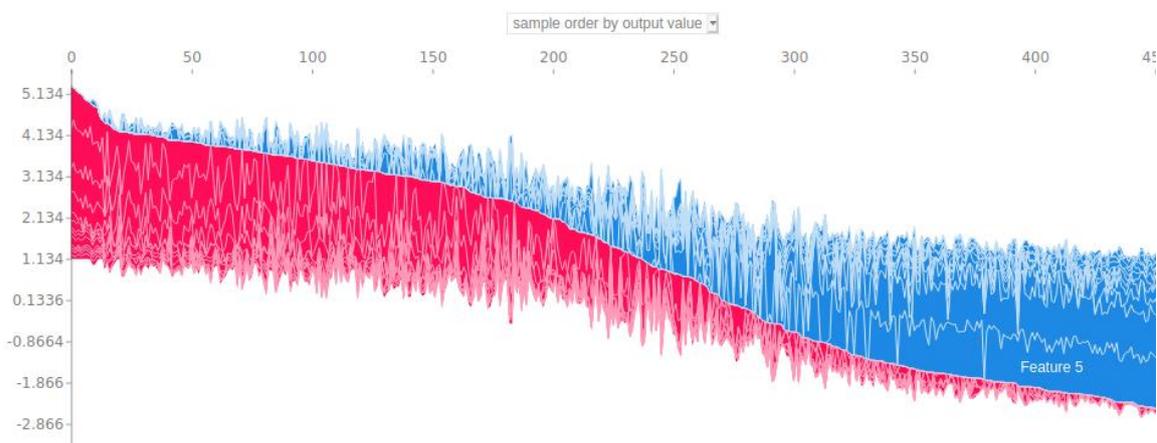

**Fig. 10.** Explain many predictions by output value

In both cases, sampling is used to implement the explanation.

It should be noted that for the evaluation of the architectures that were finally selected, thorough and long-term research was carried out on the effects of the features used in each prediction, taking into account local and universal explanations from the Shapley values methodology.

The resulting architectures are presented in the table below with the corresponding categorization performance metrics. The values in parentheses indicate the size and depth of the reservoirs, eg the Reservoir model (11-17-09) that performed and the highest results have a depth of 3, ie it includes 3 reservoirs, which include 11, 17, and 09 neurons respectively.

**Table 6.** Classification Performance Metrics of the Proposed Reservoir Models

| Classifier | Accuracy | AUC | Recall | Precision | F1 | Kappa | MCC | TT (Sec) |
|---|---|---|---|---|---|---|---|---|
| Reservoir (11-17-09) | 0.8392 | 0.9904 | 0.7502 | 0.8910 | 0.9004 | 0.8973 | 0.8807 | 59.35 |
| Reservoir (06-11-05-12) | 0.7961 | 0.9892 | 0.7483 | 0.8884 | 0.8798 | 0.8657 | 0.8623 | 68.16 |
| Reservoir (15-10-06-08) | 0.7504 | 0.9879 | 0.7469 | 0.8534 | 0.8419 | 0.8479 | 0.8588 | 53.22 |
| Reservoir (21-18-13) | 0.7492 | 0.9809 | 0.7450 | 0.8511 | 0.8415 | 0.8468 | 0.8501 | 72.38 |
| Reservoir (05-12-16-07) | 0.7262 | 0.9654 | 0.7203 | 0.8397 | 0.8402 | 0.8414 | 0.8428 | 64.04 |

By attempting an evaluation of the above results it is easy to conclude that the proposed Framework is a particularly remarkable learning system that in all evaluation cases achieved remarkable results in relation to the respective competing systems, always taking into account that the proposed system competes with corresponding systems which received training, while the proposed not.

The characteristic of this is that the Reservoir model (11-17-09) that gave the highest results, surpassed the algorithms Light Gradient Boosting Machine, Ridge Classifier, Linear Discriminant Analysis, Quadratic Discriminant Analysis, Logistic Regression, Naïve Bayes, SVM - Linear Kernel, and Ada Boost. This observation can be interpreted in the non-linearity that generally characterizes neural networks, especially in the case where it is examined where the mechanism of production of input signals is non-linear. Respectively, the resistance to structural errors of neural networks and especially the sparse architecture of reservoirs, guarantees that the malfunction or destruction of a neuron or some connections is not able to significantly disrupt their function as, the information they contain is not located in a specific point but diffuse throughout the network and especially in cases where the ESP property that characterizes reservoir architectures is achieved.

Another important observation is that the method produces extremely accurate results without recurring problems of an undetermined cause because all the features in the data set in question are handled very efficiently, based on the original features selection processes performed on the basis of PPS. This resulted in the removal of insignificant features in the network, which under certain conditions can be characterized as noise, with very negative effects on the final result.

In addition, one of the main advantages gained from the results is the high reliability resulting from kappa prices (high reliability if k≥0.70) [61], which can be considered as the result of data processing that allows the retention of the most relevant data for the forthcoming forecasts.

Finally, the use of the reservoir technique in this work is related to the fact that very often in multifactorial problems of high complexity such as the one under consideration, the prediction results are multivariate, which can be attributed to the sensitivity of the correlational models in the data. The two most important advantages of this technique focus on the fact that it offers better predictability and stability, as the overall behavior of the model is less noisy while reducing the overall risk of a particularly poor choice that may result from modeling. The above view is also supported by the dispersion of the expected error, which is concentrated close to the average error value, a fact that categorically states the reliability of the system and the generalization ability that it presents.

Obviously, a major drawback of the methodology followed, and perhaps the most serious is the highly specialized and time-consuming preprocessing process that must be followed to identify the appropriate architectures that can perform satisfactorily, which adds research complexity to data analysis and explanation. models used.

## 7. Discussion and Conclusions

An extremely innovative, reliable, low-demand, and highly efficient network traffic analysis system, based on advanced computing intelligence methods, was presented in this paper. The proposed framework implements and utilizes in the best possible way the advantages of meta-learning methodology, in order to identify malfunctions or deviations of the normal mode of operation of the network traffic, which in most cases, come from cyber attacks. The proposed digital security system was tested on a complex data set that responds to specialized operating scenarios of normal and malicious network behavior.

Our motivation in this work was to explore the processes to which only neural network architectures, without prior learning, can codify solutions and model a given task. The holistic approach proposed, which automates and solves the problem of specialized use of neural network finding algorithms without the need for human experience or intervention, is a promising approach to capture an unknown underlying mapping function between input and output data in a given problem, without requiring system training.

The methodology followed to create architectural neural networks capable of solving a given problem concerns the delimitation of the search space, which includes virtually all possible architectures that can be used in a neural network based on the data used.

As it turned out, design options, such as topology search algorithms, hyperparametric features, and their variability, the field of possible mapping functions, i.e. search space, offer an indisputable solution that optimally solves the problem under consideration. It should be emphasized that the logic of the proposed automated architecture is synonymous with the ongoing research of solutions to identify unknown digital security threats.

Under this consideration, this paper proposes an innovative and highly effective Weight Agnostic Neural Networks framework for darknet traffic, big-data analysis, and network management to real-time automating the malicious intent detection process. The proposed architecture, which is first implemented and presented in the literature, creates serious conditions for even more specialized pattern recognition systems without prior training, which are capable of responding to changing environments. It is important to emphasize that the proposed method facilitates and removes complexity from the way NAS strategies work, utilizing multiple functions of specialized methods for extracting useful intermediate representations in complex neural network architectures. The initial utilization of the predictive power of the independent variables significantly reduces the computing, producing improved training stability and remarkable categorization accuracy.

Respectively, the reservoir technology used leads to remarkable forecasting results, implementing a robust forecasting model capable of responding to highly complex problems. This ability is found in the high convergence speed of the proposed architecture which is calculated by a simple array calculation, in contrast to reciprocating stochastic gradient descent models.

To prove the validity of the proposed methodology, the SHAP methodology was used which is based on the evaluation of solutions based on Shapley values. This technique provides an understanding of how the model makes decisions and what interactions take place between the features used. Also for the precise design of the specific search space of the best architectural prototypes that can optimally solve the problem, the relationships between the variables were explored and feature selection was implemented with the Predictive Power Score technique, in order to briefly measure the predictive relationships between the available data.

It should also be emphasized that this methodology with the contribution of spatial reservoirs, deals with great precision the noisy scattered data that other spectral classification methods cannot handle.

In conclusion, the paper presents a method for the development of interpretable neural networks by encoding the solution directly in the network architecture and not in the training of its weights. Compared to other learning methods, the proposed architecture is strong in changes to node inputs, which could be the foundation for a strong defense against adversarial attacks or even damaged networks.

Providing non-zero intelligence from the beginning of the implementation of a learning system could allow the use of inherent characteristics or experience, much faster than otherwise possible, as is the case with many species in the animal kingdom.

The operation scenarios that have been proposed and simulated, implement in the most realistic way the operation of modern networks and model in a realistic way the modern harmful behaviors related to cyber-attacks. This creates possibilities for a well-defined configuration of the target database model, for high-precision categorization or correlation. It should also not be overlooked that a significant innovation lies in the fact of adding artificial intelligence to digital security mechanisms, which significantly enhance the active defense mechanisms of information systems. The proposed system, and the philosophy of active security in general, significantly enhances the ways of controlling online behavior, which can be the main means of controlling advanced cyber attacks.

Proposals for the development and future improvements of this framework should focus on the automated optimization of the appropriate parameters of the method pre-training, so as to achieve an even more efficient, accurate, and faster categorization process. It would also be important to study the expansion of this system by implementing more complex architectures with Siamese Neural Networks in an environment of parallel and distributed systems [72] or over blockchain [73], [74]. Finally, an additional element that could be studied in the direction of future expansion, concerns the operation of the network with methods of self-improvement and redefining of its parameters automatically, so that it can fully automate the process of selecting architectural hyperparameters.

**Author Contributions:** Conceptualization, K.T., D.T., K.D., L.I and C.S; methodology, K.T., D.T., K.D. and C.S; validation, K.T., D.T., K.D., L.I and C.S; formal analysis, K.T., D.T., K.D. and C.S; investigation, K.T., D.T., K.D. and C.S; writing—original draft preparation, K.T., D.T., K.D, L.I. and C.S; writing—review and editing, K.T., D.T., K.D., L.I and C.S; supervision, C.S.; project administration, K.D, L.I. All authors have read and agreed to the published version of the manuscript.